\documentclass[twocolumn,superscriptaddress,amsmath, amssymb, amsfonts,preprintnumbers,aps,prd,longbibliography]{revtex4-1}

\renewcommand{\sec}[1]{\textit{#1. --- }}

\usepackage[colorlinks=true]{hyperref} 
\usepackage{orcidlink}  
\usepackage{caption}
\usepackage{subcaption}

\usepackage{tikz}
\usetikzlibrary{decorations.pathmorphing}
\usetikzlibrary{decorations.markings}
\usetikzlibrary{positioning, shapes, snakes, arrows, calc, angles}
\tikzset{
  cross/.style={path picture={ 
      \draw[black]
      (path picture bounding box.south east) -- (path picture bounding box.north west) (path picture bounding box.south west) -- (path picture bounding box.north east);
  }},
  myCircle/.style={
    circle,
    inner sep=0pt,
    text width=5.5mm,
    align=center,
    draw=black,
    fill=white
  },
  myCircleP/.style={
    circle,
    inner sep=0pt,
    line width=.8pt,
    align=center,
    draw=black,
    fill=white
    ,minimum size=2mm
  },
  myCirclePer/.style={
    circle,
    inner sep=0pt,
    line width=.8pt,
    align=center,
    draw=black,
    fill=black
    ,minimum size=.8mm
  },
  myCirclePerPer/.style={
    circle,
    inner sep=0pt,
    line width=.8pt,
    align=center,
    draw=black,
    fill=black
    ,minimum size=2.2mm
  },
  myCirclePPP/.style={
    circle,
    inner sep=0pt,
    line width=.8pt,
    align=center,
    draw=white,
    fill=white
    ,minimum size=2.3mm
  },
  myCircleCross/.style={
    circle,cross,
    inner sep=0pt,
    line width=.8pt,
    align=center,
    draw=black,
    fill=white
    ,minimum size=2.3mm
  },
  myCirclePP/.style={
    circle,
    inner sep=0pt,
    align=center,
    draw=white,
    fill=white
    ,minimum size=5mm
  }
  ,
  myCircleDashed/.style={
    circle,
    inner sep=0pt,
    align=center,
    draw=black,
    fill=white,
    dashed,
    line width=1pt
    ,minimum size=5mm
  },
  snakeArrow/.style={line width=2pt,decorate, decoration={snake,amplitude=1.8pt,markings,mark=at position .63 with {\arrow[#1]{latex}}}},
  snakeArrowT/.style={line width=2pt,decorate, decoration={snake,amplitude=1.8pt,markings,mark=at position .5 with {\arrow[#1]{latex}}}},
  wlArrow/.style={line width=1pt,decorate, decoration={snake,amplitude=1.8pt,markings,mark=at position .63 with {\arrow[#1]{latex}}}},
  wlArrow2/.style={line width=1pt,decorate, decoration={snake,amplitude=1.8pt,markings,mark=at position .8 with {\arrow[#1]{latex}}}},
  wlArrow3/.style={line width=1pt,decorate, decoration={snake,amplitude=1.8pt,markings,mark=at position .95 with {\arrow[#1]{latex}}}},
  wlWithArrow/.style={line width=1pt,postaction={decorate},decoration={markings,mark=at position .5 with {\arrow[#1]{latex}}}},
  wl/.style={line width=1pt},
  snake/.style={line width=1pt,decorate, decoration={snake,amplitude=1pt,segment length=1.5mm}},
  background/.style={line width=1pt,dotted},
  gravitonArrow/.style={line width=.8pt, -latex,decorate, decoration={snake, segment length=3.2pt,amplitude=1.8pt, pre length=.1cm, post length=.18cm}},
  graviton/.style={line width=.8pt,decorate, decoration={snake, segment length=3.2pt,amplitude=1.8pt}},
  zParticle/.style={line width=1pt,-latex}
}
\newcommand*{\icir}[1]{}

\newcommand*{\ilabP}[1]{\scalebox{.9}{$#1$}}
\newcommand{\dd}{\mathrm{d}}
\newcommand{\nn}{\nonumber}

\newcommand{\mO}{\mathcal{O}}

\newcommand{\mM}{\mathcal{M}}

\newcommand{\mT}{\mathcal{T}}
\newcommand{\mS}{\mathcal{S}}

\newcommand{\sigo}{{\sig_1}}

\newcommand{\muo}{{\mu_1}}
\newcommand{\mut}{{\mu_2}}

\newcommand{\mun}{{\mu_n}}

\newcommand{\ab}{{\alpha\beta}}
\newcommand{\rs}{{\rho\sig}}
\newcommand{\mn}{{\mu\nu}}
\newcommand{\eps}{\epsilon}
\newcommand{\sig}{\sigma}

\newcommand{\oma}{\omega}

\newcommand{\omao}{\omega_1}

\newcommand{\pat}{\partial}
\newcommand{\del}{\delta}
\newcommand{\Del}{\Delta}

\newcommand*{\gus}[1]{}
\newcommand*{\gustav}[1]{}
\newcommand*{\new}[1]{}

\newcommand{\ce}{c_{E}}
\newcommand{\cb}{c_{B}}

\newcommand{\bse}{\begin{subequations}}
\newcommand{\ese}{\end{subequations}}

\begin{document}
\preprint{HU-EP-23/61-RTG}

\title{
  Spin and Susceptibility Effects of Electromagnetic Self-Force in Effective Field Theory
}

\author{Gustav Uhre Jakobsen\,\orcidlink{0000-0001-9743-0442}}
\email{gustav.uhre.jakobsen@physik.hu-berlin.de}
\affiliation{
  Institut f\"ur Physik und IRIS Adlershof, Humboldt Universit\"at zu Berlin, Zum Gro{\ss}en Windkanal 2, 12489 Berlin, Germany
}
\affiliation{
  Max Planck Institut f\"ur Gravitationsphysik (Albert Einstein Institut), Am M\"uhlenberg 1, 14476 Potsdam, Germany
}

\begin{abstract}
  The classic Abraham-Lorentz-Dirac self-force of point-like particles is generalized within an effective field theory setup to include linear spin and susceptibility effects described perturbatively, in that setup, by effective couplings in the action.
  Electromagnetic self-interactions of the point-like particle are integrated out using the in-in supersymmetric worldline quantum field theory formalism.
  Divergences are regularized with dimensional regularization and the resulting equations of motion are in terms only of an external electromagnetic field and the particle degrees of freedom.
\end{abstract}

\maketitle

Self-force describes the fascinating phenomenon of an object being accelerated by a force generated by itself.
The well-known Abraham-Lorentz-Dirac (ALD) equation~\cite{abraham,lorentz,Dirac:1938nz,Landau:1975pou,Jackson:1998nia} describes this effect for the most basic point-like charged particles and the resulting back-reaction balances radiation of energy described by the Larmor formula.
The physical objects of interest generally have finite extent and properties such as angular momentum (spin) and dipole susceptibilities.
For spin, adequate generalizations of the Lorentz force and corresponding ALD self-force have been considered by many authors~\cite{frenkel,bhabha1,PhysRevLett.2.435,ellis,villarroel2,Teitelboim:1979px,vanHolten:1990we,Kazinski:2006dh,Kosyakov:2018wek}.
One motivation for this line of work is the classical description of the electron~\cite{KIM19991} which may e.g. be modeled as a charged sphere for which several self-force results are known~\cite{rohrlich,Galley:2010es}.

Recently, an analogous problem in gravity of describing the early inspiral of two point-like compact bodies and their radiation have gained importance for the data analysis of gravitational wave signals observed on earth~\cite{LIGOScientific:2016aoc,LIGOScientific:2021djp}.
Here, one sets up an effective field theory (EFT) capturing the body degrees of freedom by worldline fields with the most basic field given by the worldline parametrization $z^\mu(\tau)$~\cite{Goldberger:2004jt,Goldberger:2022ebt}.
Spin and finite size effects are then described by effective couplings whose value may in each case be determined from a matching to the physical object of interest.
Such a worldline EFT has had great success in describing compact bodies in gravity~\cite{Goldberger:2006bd,Porto:2016pyg,Levi:2018nxp} but may also be applied to electromagnetic interactions~\cite{Galley:2010es,Birnholtz:2014fwa,Patil:2020dme,Bhattacharyya:2023kbh,Kim:2023drc}.

In the same context of gravitational wave physics, quantum field theoretic methods have been used advantageously to describe classical physics~\cite{Neill:2013wsa,Bjerrum-Bohr:2018xdl,Cheung:2018wkq,Bern:2019nnu}.
In this spirit, classical dynamics as described by worldline EFT may be considered as the tree-level contributions of a worldline quantum field theory (WQFT)~\cite{Mogull:2020sak,Jakobsen:2021smu,Jakobsen:2021lvp,Jakobsen:2021zvh,Jakobsen:2022fcj,Jakobsen:2022psy,Jakobsen:2022zsx,Jakobsen:2023ndj,Jakobsen:2023ndj,Jakobsen:2023oowPP,Jakobsen:2023hig}.
This gives rise to an efficient diagrammatic approach to solving the classical equations of motion.
In this action-based framework, causal boundary conditions are imposed with the Schwinger-Keldysh in-in prescription~\cite{Schwinger:1960qe,Keldysh:1964ud,Galley:2008ih,Galley:2009px,Galley:2012hx,Birnholtz:2013nta,Polonyi:2013lma,Almeida:2022jrv,Kalin:2022hph,Jakobsen:2022psy}.
Several state of the art results in the perturbative expansion of gravitational scattering have been computed with the WQFT~\cite{Jakobsen:2021lvp,Jakobsen:2022fcj,Jakobsen:2022zsx,Jakobsen:2023ndj,Jakobsen:2023hig} (see also~\cite{Shi:2021qsb,Comberiati:2022cpm,Bastianelli:2021nbs,Comberiati:2022ldk,Wang:2022ntx,Kim:2023drc,Ben-Shahar:2023djm} for additional work with WQFT).

In worldline EFT, the relativistic angular momentum of the point-like particle is described by an antisymmetric worldline tensor field $S^\mn(\tau)$.
Half of its degrees of freedom are constrained by requiring symmetry of the action under small shifts of the worldline trajectory~\cite{Steinhoff:2015ksa} so that the dynamics involves only a spatial spin vector.
At the level of the action, one must usually introduce a co-moving frame in order to describe the spin kinematics~\cite{Porto:2005ac,Levi:2015msa,Ben-Shahar:2023djm}.
This, however, is avoided by expressing the spin tensor in terms of anticommuting Grassmann vectors $\psi^\mu(\tau)$ which, inspired by previous work~\cite{Berezin:1976eg,Brink:1976sz,Brink:1976uf,Barducci:1976wc,Galvao:1980cu,Howe:1988ft,Howe:1989vn,vanHolten:1990we,Gibbons:1993ap,Bastianelli:2005vk,Bastianelli:2005uy}, was first proposed in this context in the framework of WQFT~\cite{Jakobsen:2021lvp,Jakobsen:2021zvh}.
Here, the worldline shift symmetry becomes a supersymmetry (SUSY).

Self-interaction of point-like particles generally leads to divergent expressions which, however, from the perspective of EFT is not surprising as the small scale physics has been integrated out.
Instead, the EFT must be regularized and in the present case we will use dimensional regularization.
Thus, also in the classical setting, eventual divergences must be absorbed into counterterms of the action~\cite{Goldberger:2004jt,Galley:2015kus,Jakobsen:2020ksu,Mandal:2023hqa,Barack:2023oqp}.

In this letter, we compute novel spin and susceptibility corrections to the electromagnetic self-force of point-like particles described by a worldline EFT.
The computational method innovates on earlier work and presents a very streamlined approach for deriving self-force corrections in worldline EFT.
In particular, computations are carried out diagrammatically using the in-in SUSY WQFT formalism and reduce to the evaluation of a number of tree-level Feynman diagrams.
A major motivation for this innovation is its future generalization and application to the gravitational setting and, in particular, the perturbative self-force expansion of extreme mass ratio binaries~\cite{Mino:1996nk,Galley:2008ih,Poisson:2011nh,Barack:2018yvs}.

\sec{EFT of Point-Like Particles}
Our system will be described by the following action $S$:
\begin{equation}
  \label{TotalAction}
  S=S_{\rm kin}+S_{\rm int}+S_{\rm EM}+S_{\rm ext}
  \ .
\end{equation}
The first two terms will describe kinematics and electromagnetic (EM) interactions of the point-like particle.
The third term is the kinetic action of the EM potential in Lorenz gauge,
\begin{equation}
  \label{EMBulk}
  S_{\rm EM}
  =
  -\!
  \int \dd^d x
  \Big[
  \frac14
  F^\mn(x) F_\mn(x)
  +
  \frac12
  \big[
    \pat_\mu A^\mu(x)
    \big]^2
  \Big]
  \, ,
\end{equation}
with arbitrary dimension $d$ for the use of dimensional regularization and field strength tensor $F_\mn=2\pat_{[\mu} A_{\nu]}$ where square brackets denote averaged antisymmetrization.
We use units such that the speed of light and vacuum permittivity and permeability are all unity $c=\eps_0=\mu_0=1$.
Finally, the last term of Eq.~\eqref{TotalAction}, $S_{\rm ext}$, describes external sources of the EM potential.
We do not make any assumptions on $S_{\rm ext}$ which could for example be given by a second copy of the worldline action in which case we would describe the relativistic EM two-body problem.

Let us first consider the interaction terms of the point-like particle which we model as follows:
\begin{eqnarray}
  S_{\rm int}
  \!\!&=&\!\!
  -\int\dd\tau
  \Big(
  q\,\dot z
  \cdot
  A(z)
  -
    \frac{q}{m}
    \dot z_\mu
    \mS^\mn
    E_\nu(z)
    +
    |\dot z|
  U
  \Big)
  \ ,
  \nn
  \\
  U
  \!\!&=&\!\!
  \frac{gq}{2m}
    S
    \cdot
    B(z)
    +
    \frac{\cb}2
    B^2(z)
    +
    \frac{\ce
    }{2}
    E^2(z)
    \ .
    \label{interactions}
\end{eqnarray}
Here, $z^\mu=z^\mu(\tau)$ is the worldline of the point-like particle with total charge $q$ and mass $m$ and we use dots to denote differentiation with respect to $\tau$ and the shorthand $|\dot z| = \sqrt{\dot z^2}$ with factors of $|\dot z|$ ensuring explicit time reparametrization invariance.
The particle has (intrinsic) relativistic angular momentum $\mS^\mn(\tau)$ with Pauli-Lubanski vector $S^\mu=\frac12\eps^{\mn\rs}\mS_{\nu\rho}\dot z_\sig/|\dot z|$.
The electric and magnetic fields $E^\mu(z)$ and $B^\mu(z)$ are defined implicitly by a decomposition of the field strength tensor $F^\mn(z)$,
\begin{equation}
    \label{EBFields}
  F^\mn(z)
  =
  \frac1{|\dot z|}
  \Big(
  2 E^{[\mu}(z)
    \dot z^{\nu]}
  +
  \eps^{\mn}_{\ \ \rs}
  B^\rho(z)
  \dot z^\sig
  \Big)
  \ ,
\end{equation}
where the vectors are assumed to be orthogonal to the body frame ($B\cdot\dot z=E\cdot\dot z=0$).
Here, and in the following, we often leave time dependence of worldline fields implicit.

In Eq.~\eqref{interactions}, the spin-induced magnetic field is measured by the $g$-factor $g$ and susceptibility effects by $\cb$ and $\ce$ describing magnetization and electric polarization, respectively.
The interactions Eq.~\eqref{interactions} are invariant (at leading order in spin and susceptibility) under small shifts of the trajectory $\del z^\mu$ where the spin tensor transforms as $\delta \mS^\mn=2m \delta z^{[\mu}\dot z^{\nu]}/|\dot z|$ and the Pauli-Lubanski vector is invariant.
Here, and after, for the use of dimensional regularization, all Levi-Civita symbols may be avoided by working with $\mS^\mn$ and $F^\mn$ as discussed explicitly in the supplementary material~\cite{supp}.

If one assumes $L\sim q^2/m$ to be the only scale of the point-like particle, one finds $S^\mu\sim L m$ and $c_{\rm E/B}\sim L^3$ although, generally, additional intrinsic scales may be relevant.
The EFT framework assumes this scale to be small compared with a relevant external scale and effective couplings are further suppressed by it.
The inclusion of higher order spin or susceptibility corrections or other finite size effects in the EFT is an interesting problem with much work done in the gravitational context~\cite{Levi:2015msa,Saketh:2022wap,Saketh:2022xjb,Haddad:2023ylx,Aoude:2023vdk}.

Let us turn to the kinetic action $S_{\rm kin}$ which, as discussed in the introduction, is conveniently written in terms of anticommuting (Hermitian) Grassmann vectors $\psi^\mu(\tau)$ related to the the spin tensor as $\mS^\mn=-im\psi^\mu\psi^\nu$.
Using also the Polyakov form of the point mass action, we get~\cite{Jakobsen:2021zvh,Jakobsen:2022zsx,Jakobsen:2023oowPP}:
\begin{equation}
  S_{\rm kin}
  =
  -\frac{m}2
  \int\dd\tau
  \Big(
  \dot z^2
  +
  i\psi\cdot \dot \psi\Big)
  \ .
\end{equation}
At this point, the shift symmetry becomes a SUSY with $\delta z^\mu=i\eta \psi^\mu$ and $\delta \psi^\mu = -\eta \dot z^\mu$ and global Grassmann parameter $\eta$.
We will gauge-fix the SUSY with the covariant spin supplementary condition $S^\mn \dot z_\nu=0$ and time reparametrization invariance with proper time $\dot z^2=1$ and assume these constraints in the following.

\sec{Worldline Equations of Motion}
The equations of motion (EOMs) are derived from the principle of stationary action and for the trajectory we find the force $f^\sig=m\ddot z^\sig$ to be:
\begin{eqnarray}
      \label{tForce}
  f^\mu
  \!\!&=&\!\!
  q E^\mu(z)
  +
  (\eta_\bot^\mn
  \pat_\nu
  -\ddot z^\mu)
  U
  \\
  &&\!\!-
  \eta_\bot^\mn
  \frac{\dd}{\dd\tau}
  \Big[
    \Big(
    \frac{(g-2) q}{2m}
    S
    +
    \big(
    \ce+\cb
    \big)
    B(z)
    \Big)
    \times
    E(z)
    \Big]_\nu 
    \,.
  \nn
\end{eqnarray}
Here, we use a projector $\eta_\bot^\mn=\eta^\mn- \dot z^\mu \dot z^\nu$ and note that proper time implies $\dot z\cdot f=0$.
We define the body frame cross product of any two vectors $u_1^\mu$ and $u_2^\mu$ by
\begin{equation}\label{crossP}
  (u_1\times u_2)^\mu
  =
  \eps^\mu_{\ \nu\rs}
  u_1^\nu
  u_2^\rho
  \dot z^\sigma
  \ ,
\end{equation}
which implies $\eps^{1230}=1$.

We will focus on the (SUSY invariant) Pauli-Lubanski vector $S^\mu(\tau)$ as the physical spin variable which is given in terms of the Grassmann vectors by $S^\mu=-i\frac{m}{2}(\psi\times\psi)^\mu$.
Using the chain rule and principle of stationary action for the Grassmann vectors one arrives at the following spin precession for $S^\mu$ (the BMT equation~\cite{PhysRevLett.2.435,Kim:2023drc}):
\begin{equation}
  \label{torque}
  \eta_{\bot\nu}^\mu\dot S^\nu
  =
  \mT^\mu
  =
  \frac{gq}{2m}
  \big(
  S\times B(z)
  \big)^\mu
  \ .
\end{equation}
Here, we introduced the torque $\mT^\mu$ and focused only on the spatial components as the time component of $\dot S^\mu$ in the direction of $\dot z^\mu$ is straightforwardly determined from differentiation of the constraint $S\cdot \dot z=0$.

\sec{Worldline Quantum Field Theory}
The WQFT formalism offers a streamlined diagrammatic approach to solving the classical EOMs~\eqref{tForce} and~\eqref{torque}~\cite{Mogull:2020sak,Jakobsen:2021smu,Jakobsen:2021lvp,Jakobsen:2021zvh,Jakobsen:2022fcj,Jakobsen:2022psy,Jakobsen:2022zsx,Jakobsen:2023ndj,Jakobsen:2023ndj,Jakobsen:2023oowPP,Jakobsen:2023hig}.
The central idea is that the classical dynamics described by the worldline EFT may be considered as the tree-level contributions ($\hbar\to0$) of a quantum field theory defined from the (worldline) action $S$ where both the EM potential and the worldline fields are promoted to quantum fluctuating fields.

For the EM potential, we define the fluctuating field $\Del A^\mu=A^\mu-A_{\rm ext}^\mu$ in a background expansion around the external potential $A_{\rm ext}^\mu(x)$ sourced by the current of $S_{\rm ext}$ such that
\begin{equation}\label{externalField}
  \pat^2 A_{\rm ext}^\mu(x)
  =
  -\frac{\del S_{\rm ext}}{\del A_\mu(x)}
  \ .
\end{equation}
We collect the worldline fields in a single superfield $Z^\mu=\{z^\mu,\psi^\mu\}$ and expand it around an arbitrary time $\bar \tau$,
\begin{equation}
  Z^\sig(\tau)
  =
  \big\{
  z^\sig(\bar\tau)+
  \big(
  \tau-\bar\tau
  \big)
  \dot z^\sig(\bar\tau)
  ,
  \psi^\sig(\bar\tau)
  \big\}
  +
  \Del Z^\sig(\tau)
  \ ,
\end{equation}
with fluctuation $\Del Z^\sig=\{\Del z^\sig,\Del \psi^\sig\}$ and boundary conditions $\Del z(\bar\tau)=\Del\dot z(\bar\tau)=\Del \psi(\bar\tau)=0$.
This expansion will be used at times near to $\bar\tau$ and implies no assumptions on the global character of the trajectory.

The key observation of the WQFT formalism is that its one-point functions in the $\hbar\to0$ limit are identical to the solutions of the classical EOMs:
\begin{equation}
  \Del A^\mu(k)
  =
   \begin{tikzpicture}[baseline={([yshift=-.5ex]u1)}]
    \def\upN{0} 
    \def\cN{0} 
    \def\laN{.7} 
    \coordinate (u1) at (\cN,\upN) ;
    \coordinate (u2) at (\cN+\laN,\upN) ;
    \draw[gravitonArrow] (u2) -- (u1) ;
    \node[myCircleP] () at (u2) {\icir{S}} ;
   \end{tikzpicture}
   \ ,
   \qquad
   \Del Z^\sig(\oma)
   =
   \begin{tikzpicture}[baseline={([yshift=-.5ex]u1)}]
    \def\upN{0} 
    \def\cN{0} 
    \def\laN{.7} 
    \coordinate (u1) at (\cN,\upN) ;
    \coordinate (u2) at (\cN+\laN,\upN) ;
    \draw[zParticle] (u2) -- (u1) ;
    \node[myCircleP] () at (u2) {\icir{S}} ;
     \end{tikzpicture}
     \ .
\end{equation}
Here, the blobs represent the WQFT one-point functions with wiggly and solid lines identifying photons $\Del A^\mu$ and superfields respectively.
Conveniently, we work in momentum and frequency space indicated by $k^\mu$ and $\oma$ and defined by $d$-dimensional and one-dimensional Fourier transforms, respectively.
In order to consider arbitrary times $\tau$, the corresponding frequency must be kept off-shell (the on-shell limit $\oma\to0$ is related to the global change in momentum or spin~\cite{Mogull:2020sak,Jakobsen:2023oowPP}).

The WQFT Feynman rules are straightforwardly determined from the action~\cite{Mogull:2020sak,Jakobsen:2021zvh,Jakobsen:2023oowPP} and have the following three important properties.
First, the background expansion introduces one-point vertices which lead to an infinite series of tree diagrams.
Second, the interaction of one-dimensional superfields with $d$-dimensional photons conserves only one component of the photon momenta and the unconstrained integration on the remaining (spatial) components leads to loop-like integrations within the tree diagrams.
Third, in order to arrive at causal dynamics, retarded propagators are used exclusively and all point toward the single outgoing line which, formally, is imposed by the in-in formalism~\cite{Jakobsen:2022psy}.

A simple example of a vertex rule is given by the interaction of a photon with a worldline trajectory fluctuation,
\begin{equation}
  \label{feynmanRules1}
  \begin{tikzpicture}[baseline={([yshift=-.5ex]u0)}]
     \coordinate (u0) at (-.65,0) ;
     \coordinate (u2) at (-.3,0) ;
     \coordinate (u1) at (0,0) ;
     \coordinate (u3) at (.65,0) ;
     \coordinate (u4) at (.3,-.05) ;
     \draw[graviton] (u1) -- (u0) ;
     \draw[wl] (u3) -- (u1) ;
     \node[myCirclePer] () at (u1) {} ;
     \node[below] () at (u2) {
       \ilabP{\Del A_{\mu}(-k)\ \ \ }} ;
     \node[above] () at (u4) {
       \ilabP{\ \ \ \Del z^\sig(\oma)}} ;
  \end{tikzpicture}
  \hspace{-.2cm}
  =\!
  4\pi\delta
  \big(
  k\cdot \dot z
  -
  \oma
  \big)
  e^{i k\cdot(z-\tau\dot z)}
  k^{[\sig}
    \eta^{\nu]}_\mu
  \dot z_\nu
  +...
  \Big|_{\tau\to\bar\tau}
  \,,
\end{equation}
with the ellipsis indicating spin and susceptibility corrections.
Generally, the vertex rules have up to two photon legs and any number of superfield legs.
They conserve energy and depend on the worldline background variables $z^\sig(\bar\tau)$, $\dot z^\sig(\bar\tau)$ and $\psi^\sig(\bar\tau)$, on the external EM potential $A_{\rm ext}^\mu$ and on the momenta and frequencies of the incoming and outgoing fields.
Because of the background expansion around $A_{\rm ext}^\mu$, the photons $\Del A^\mu(x)$ interact only with the point-like particle (and not the external current).

The classical EOMs now take the form of off-shell, recursive Berends-Giele like relations~\cite{Jakobsen:2023ndj,Jakobsen:2023oowPP,BERENDS1988759}:
\begin{equation}
  \label{BGZ}
  \begin{aligned}
    \begin{tikzpicture}[baseline={([yshift=-.5ex]u1)}]
      \def\upN{0} 
      \def\cN{0} 
      \def\laN{.5} 
      \coordinate (u1) at (\cN,\upN) ;
      \coordinate (u2) at (\cN+\laN,\upN) ;
      \draw[zParticle] (u2) -- (u1) ;
      \node[myCircleP] () at (u2) {\icir{S}} ;
    \end{tikzpicture}
    &=
    \!\sum_{n=0}^{\infty}
    \frac{1}{n!}
    \bigg[
      \begin{tikzpicture}[baseline={([yshift=-.5ex]u1)}]
        \def\upN{0} 
        \def\cN{0} 
        \def\downNa{-.4} 
        \def\downNb{-.4} 
        \def\laN{.5} 
        \def\lbN{.4} 
        \def\lcN{.4} 
        \def\ldN{.4} 
        \def\angleN{90} 
        \def\leN{-1.0} 
        \def\lfN{.4} 
        \def\correc{.4} 
        \def\degA{17}
        \def\lbA{.5}
        \def\lcA{.3}
        \def\ldA{.4}
        \coordinate (u1) at (\cN,\upN) ;
        \coordinate (d1) at (\cN,\upN+\leN) ;
        \coordinate (u2) at (\cN+\laN,\upN) ;
        \coordinate (ur2) at (\cN+\laN+\lbA,-\upN+\lcA) ;
        \coordinate (ul2) at (\cN+\laN+\lbA,-\upN-\lcA) ;
        \coordinate (d2) at (\cN+\laN,\upN+\leN) ;
        \coordinate (dr2) at (\cN+\laN+\lfN,\upN+\leN) ;
        \coordinate (dl2) at (\cN+\laN-\lfN,\upN+\leN) ;
        \draw[zParticle] (u2) -- (u1) ;
        \draw[wl] (u2) -- (ur2) ;
        \draw[wl] (u2) -- (ul2) ;
        \draw[background] (u2) ++(\degA:\ldA) arc (\degA:-\degA:\ldA) ;
        \node[right=.36cm] () at (u2) {$n$} ;
        \node[myCirclePer] () at (u2) {\icir{S}} ;
        \node[myCircleP] () at (ur2) {\icir{S}} ;
        \node[myCircleP] () at (ul2) {\icir{S}} ;
      \end{tikzpicture}
      +
      \begin{tikzpicture}[baseline={([yshift=-.5ex]u1)}]
        \def\upN{0} 
        \def\cN{0} 
        \def\downNa{-.4} 
        \def\downNb{-.4} 
        \def\laN{.5} 
        \def\lbN{.4} 
        \def\lcN{.4} 
        \def\ldN{.4} 
        \def\angleN{90} 
        \def\leN{.5} 
        \def\lfN{.4} 
        \def\correc{.4} 
        \def\degA{17}
        \def\lbA{.5}
        \def\lcA{.3}
        \def\ldA{.4}
        \coordinate (u1) at (\cN,\upN) ;
        \coordinate (d1) at (\cN,\upN+\leN) ;
        \coordinate (u2) at (\cN+\laN,\upN) ;
        \coordinate (ur2) at (\cN+\laN+\lbA,-\upN+\lcA) ;
        \coordinate (ul2) at (\cN+\laN+\lbA,-\upN-\lcA) ;
        \coordinate (d2) at (\cN+\laN,\upN+\leN) ;
        \coordinate (dr2) at (\cN+\laN+\lfN,\upN+\leN) ;
        \coordinate (dl2) at (\cN+\laN-\lfN,\upN+\leN) ;
        \coordinate (u3) at (\cN+\laN+\lbN,\upN) ;
        \coordinate (d3) at (\cN+\laN+\lbN,\upN+\leN) ;
        \coordinate (u4) at (\cN+\laN+\lbN+\lcN,\upN) ;
        \coordinate (d4) at (\cN+\laN+\lbN+\lcN,\upN+\leN) ;
        \coordinate (u5) at (\cN+\laN+\lbN+\lcN+\ldN,\upN) ;
        \coordinate (d5) at (\cN+\laN+\lbN+\lcN+\ldN,\upN+\leN) ;
        \coordinate (mm1) at (\cN+\laN+\lbN/2,\upN+\downNa) ;
        \coordinate (mm2) at (\cN+\laN+\lbN+\lcN/2,\upN+\downNb) ;
        \coordinate (arrowL) at (-.2,0) ;
        \coordinate (arrowH) at (0,+.2) ;
        \coordinate (u2P) at ($(u2)-(\cN+\correc,\upN)$) ;
        \draw[zParticle] (u2) -- (u1) ;
        \draw[graviton] (d2) -- (u2) ;
        \draw[wl] (u2) -- (ur2) ;
        \draw[wl] (u2) -- (ul2) ;
        \draw[background] (u2) ++(\degA:\ldA) arc (\degA:-\degA:\ldA) ;
        \node[right=.36cm] () at (u2) {$n$} ;
        \node[myCirclePer] () at (u2) {\icir{S}} ;
        \node[myCircleP] () at (ur2) {\icir{S}} ;
        \node[myCircleP] () at (ul2) {\icir{S}} ;
        \node[myCircleP] () at (d2) {\icir{S}} ;
      \end{tikzpicture}
      +
      \frac12
      \begin{tikzpicture}[baseline={([yshift=-.5ex]u1)}]
        \def\upN{0} 
        \def\cN{0} 
        \def\laN{.5} 
        \def\lbN{.4} 
        \def\lcN{.4} 
        \def\angleN{90} 
        \def\leN{.5} 
        \def\lfN{.2} 
        \def\degA{17}
        \def\lbA{.5}
        \def\lcA{.3}
        \def\ldA{.4}
        \coordinate (u1) at (\cN,\upN) ;
        \coordinate (d1) at (\cN,\upN+\leN) ;
        \coordinate (u2) at (\cN+\laN,\upN) ;
        \coordinate (ur2) at (\cN+\laN+\lbA,-\upN+\lcA) ;
        \coordinate (ul2) at (\cN+\laN+\lbA,-\upN-\lcA) ;
        \coordinate (d2) at (\cN+\laN,\upN+\leN) ;
        \coordinate (dr2) at (\cN+\laN+\lfN,\upN+\leN) ;
        \coordinate (dl2) at (\cN+\laN-\lfN,\upN+\leN) ;
        \draw[zParticle] (u2) -- (u1) ;
        \draw[graviton] (dr2) -- (u2) ;
        \draw[graviton] (dl2) -- (u2) ;
        \draw[wl] (u2) -- (ur2) ;
        \draw[wl] (u2) -- (ul2) ;
        \draw[background] (u2) ++(\degA:\ldA) arc (\degA:-\degA:\ldA) ;
        \node[right=.36cm] () at (u2) {$n$} ;
        \node[myCirclePer] () at (u2) {\icir{S}} ;
        \node[myCircleP] () at (ur2) {\icir{S}} ;
        \node[myCircleP] () at (ul2) {\icir{S}} ;
        \node[myCircleP] () at (dr2) {\icir{S}} ;
        \node[myCircleP] () at (dl2) {\icir{S}} ;
      \end{tikzpicture}
      \bigg]
    \, ,
    \\
    \begin{tikzpicture}[baseline={([yshift=-.5ex]u1)}]
      \def\upN{0} 
      \def\cN{0} 
      \def\laN{.5} 
      \coordinate (u1) at (\cN,\upN) ;
      \coordinate (u2) at (\cN+\laN,\upN) ;
      \draw[gravitonArrow] (u2) -- (u1) ;
      \node[myCircleP] () at (u2) {} ;
    \end{tikzpicture}
    &=
    \!\sum_{n=0}^{\infty}
    \frac{1}{n!}
    \bigg[
      \begin{tikzpicture}[baseline={([yshift=-.5ex]u1)}]
        \def\upN{0} 
        \def\cN{0} 
        \def\downNa{-.4} 
        \def\downNb{-.4} 
        \def\laN{.6} 
        \def\lbN{.4} 
        \def\lcN{.4} 
        \def\ldN{.4} 
        \def\angleN{90} 
        \def\leN{-1.0} 
        \def\lfN{.4} 
        \def\correc{.4} 
        \def\degA{17}
        \def\lbA{.5}
        \def\lcA{.3}
        \def\ldA{.4}
        \coordinate (u1) at (\cN,\upN) ;
        \coordinate (d1) at (\cN,\upN+\leN) ;
        \coordinate (u2) at (\cN+\laN,\upN) ;
        \coordinate (ur2) at (\cN+\laN+\lbA,-\upN+\lcA) ;
        \coordinate (ul2) at (\cN+\laN+\lbA,-\upN-\lcA) ;
        \coordinate (d2) at (\cN+\laN,\upN+\leN) ;
        \draw[gravitonArrow] (u2) -- (u1) ;
        \draw[wl] (u2) -- (ur2) ;
        \draw[wl] (u2) -- (ul2) ;
        \draw[background] (u2) ++(\degA:\ldA) arc (\degA:-\degA:\ldA) ;
        \node[right=.36cm] () at (u2) {$n$} ;
        \node[myCirclePer] () at (u2) {\icir{S}} ;
        \node[myCircleP] () at (ur2) {\icir{S}} ;
        \node[myCircleP] () at (ul2) {\icir{S}} ;
      \end{tikzpicture}
      +
      \begin{tikzpicture}[baseline={([yshift=-.5ex]u1)}]
        \def\upN{0} 
        \def\cN{0} 
        \def\downNa{-.4} 
        \def\downNb{-.4} 
        \def\laN{.6} 
        \def\lbN{.4} 
        \def\lcN{.4} 
        \def\ldN{.4} 
        \def\angleN{90} 
        \def\leN{.5} 
        \def\lfN{.4} 
        \def\degA{17}
        \def\lbA{.5}
        \def\lcA{.3}
        \def\ldA{.4}
        \coordinate (u1) at (\cN,\upN) ;
        \coordinate (d1) at (\cN,\upN+\leN) ;
        \coordinate (u2) at (\cN+\laN,\upN) ;
        \coordinate (ur2) at (\cN+\laN+\lbA,-\upN+\lcA) ;
        \coordinate (ul2) at (\cN+\laN+\lbA,-\upN-\lcA) ;
        \coordinate (d2) at (\cN+\laN,\upN+\leN) ;
        \draw[gravitonArrow] (u2) -- (u1) ;
        \draw[graviton] (d2) -- (u2) ;
        \draw[wl] (u2) -- (ur2) ;
        \draw[wl] (u2) -- (ul2) ;
        \draw[background] (u2) ++(\degA:\ldA) arc (\degA:-\degA:\ldA) ;
        \node[right=.36cm] () at (u2) {$n$} ;
        \node[myCirclePer] () at (u2) {\icir{S}} ;
        \node[myCircleP] () at (ur2) {\icir{S}} ;
        \node[myCircleP] () at (ul2) {\icir{S}} ;
        \node[myCircleP] () at (d2) {\icir{S}} ;
      \end{tikzpicture}
      \bigg]
    \, .
  \end{aligned}
\end{equation}
The first line corresponds to the worldline EOMs where the first term represents the force (or torque) evaluated on the external EM fields and the next two terms have one or two insertions of the fluctuation $\Del A^\mu(x)$.
This force is expanded in the worldline fluctuations around the background time $\bar\tau$ which explains the presence of any number $n$ of fluctuations.
When evaluated at the background time itself in time domain, only finitely many terms in the sum on $n$ are non-zero.
Such an evaluation at $\bar\tau$ will be our goal after integrating out $\Del A^\mu(x)$ below.
The second line of Eq.~\eqref{BGZ} describes the coupling of $\Del A^\mu$ to the current of the point-like particle.

\sec{Integrating Out Self-Interactions}
Self-interactions are now straightforwardly integrated out by eliminating $\Del A^\mu(x)$ from the system of equations~\eqref{BGZ} leading to the following regulated EOM:
\begin{widetext}
\begin{equation}
  \label{exp1}
  \begin{tikzpicture}[baseline={([yshift=-.5ex]u1)}]
    \def\upN{0} 
    \def\cN{0} 
    \def\laN{.5} 
    \coordinate (u1) at (\cN,\upN) ;
    \coordinate (u2) at (\cN+\laN,\upN) ;
    \draw[zParticle] (u2) -- (u1) ;
    \node[myCircleP] () at (u2) {\icir{S}} ;
  \end{tikzpicture}
  =
  \sum
  \Bigg[
  \frac1{n!}
    \begin{tikzpicture}[baseline={([yshift=-.5ex]u1)}]
      \def\upN{0} 
      \def\cN{0} 
      \def\laN{.5} 
      \def\degA{18}
      \def\lbA{.3}
      \def\lcA{.5}
      \def\ldA{.4}
      \coordinate (u1) at (\cN,\upN) ;
      \coordinate (u2) at (\cN+\laN,\upN) ;
      \coordinate (ur2) at (\cN+\laN+\lbA,-\upN-\lcA) ;
      \coordinate (ul2) at (\cN+\laN-\lbA,-\upN-\lcA) ;
      \draw[zParticle] (u2) -- (u1) ;
      \draw[wl] (u2) -- (ur2) ;
      \draw[wl] (u2) -- (ul2) ;
      \draw[background] (u2) ++(\degA-90:\ldA) arc (\degA-90:-\degA-90:\ldA) ;
      \node[myCirclePer] () at (u2) {\icir{S}} ;
      \node[myCircleP] () at (ur2) {} ;
      \node[myCircleP] () at (ul2) {} ;
      \node[below=.36cm] () at (u2) {$n$} ;
    \end{tikzpicture}
    +
    \frac1{n!m!}
    \begin{tikzpicture}[baseline={([yshift=-.5ex]u1)}]
      \def\upN{0} 
      \def\cN{0} 
      \def\downNa{.4} 
      \def\laN{.5} 
      \def\lbN{.9} 
      \def\lcN{.4} 
      \def\ldN{.6} 
      \def\angleN{-90} 
      \def\leN{-1.0} 
      \def\degA{18}
      \def\lbA{.3}
      \def\lcA{-.5}
      \def\ldA{.4}
      \coordinate (u1) at (\cN,\upN) ;
      \coordinate (u2) at (\cN+\laN,\upN) ;
      \coordinate (ur2) at (\cN+\laN+\lbA,\upN+\lcA) ;
      \coordinate (ul2) at (\cN+\laN-\lbA,\upN+\lcA) ;
      \coordinate (u3) at (\cN+\laN+\lbN,\upN) ;
      \coordinate (ur3) at (\cN+\laN+\lbN+\lbA,\upN+\lcA) ;
      \coordinate (ul3) at (\cN+\laN+\lbN-\lbA,\upN+\lcA) ;
      \coordinate (mm1) at (\cN+\laN+\lbN/2,\upN+\downNa) ;
      \draw[zParticle] (u2) -- (u1) ;
      \draw[graviton] (u3) to[out=-180+\angleN,in=0] (mm1) to[out=-180,in=-\angleN] (u2);
      \draw[wl] (u2) -- (ur2) ;
      \draw[wl] (u2) -- (ul2) ;
      \draw[background] (u2) ++(\degA-90:\ldA) arc (\degA-90:-\degA-90:\ldA) ;
      \draw[wl] (u3) -- (ur3) ;
      \draw[wl] (u3) -- (ul3) ;
      \draw[background] (u3) ++(\degA-90:\ldA) arc (\degA-90:-\degA-90:\ldA) ;
      \node[myCirclePer] () at (u2) {\icir{S}} ;
      \node[myCirclePer] () at (u3) {\icir{S}} ;
      \node[myCircleP] () at (ur2) {} ;
      \node[myCircleP] () at (ul2) {} ;
      \node[below=.36cm] () at (u2) {$n$} ;
      \node[myCircleP] () at (ur3) {} ;
      \node[myCircleP] () at (ul3) {} ;
      \node[below=.36cm] () at (u3) {$m$} ;
    \end{tikzpicture}
    +
    \frac1{n!m!l!}
    \begin{tikzpicture}[baseline={([yshift=-.5ex]u1)}]
      \def\upN{0} 
      \def\cN{0} 
      \def\downNa{.4} 
      \def\downNb{.4} 
      \def\laN{.5} 
      \def\lbN{.9} 
      \def\lcN{.9} 
      \def\ldN{.4} 
      \def\angleN{-90} 
      \def\leN{-1.0} 
      \def\degA{18}
      \def\lbA{.3}
      \def\lcA{-.5}
      \def\ldA{.4}
      \coordinate (u1) at (\cN,\upN) ;
      \coordinate (u2) at (\cN+\laN,\upN) ;
      \coordinate (d2) at (\cN+\laN,\upN+\leN) ;
      \coordinate (u3) at (\cN+\laN+\lbN,\upN) ;
      \coordinate (u4) at (\cN+\laN+\lbN+\lcN,\upN) ;
      \coordinate (mm1) at (\cN+\laN+\lbN/2,\upN+\downNa) ;
      \coordinate (mm2) at (\cN+\laN+\lbN+\lcN/2,\upN+\downNb) ;
      \coordinate (arrowL) at (-.2,0) ;
      \coordinate (ur2) at (\cN+\laN+\lbA,\upN+\lcA) ;
      \coordinate (ul2) at (\cN+\laN-\lbA,\upN+\lcA) ;
      \coordinate (ur3) at (\cN+\laN+\lbN+\lbA,\upN+\lcA) ;
      \coordinate (ul3) at (\cN+\laN+\lbN-\lbA,\upN+\lcA) ;
      \coordinate (ur4) at (\cN+\laN+\lbN+\lcN+\lbA,\upN+\lcA) ;
      \coordinate (ul4) at (\cN+\laN+\lbN+\lcN-\lbA,\upN+\lcA) ;
      \draw[zParticle] (u2) -- (u1) ;
      \draw[graviton] (u3) to[out=-180+\angleN,in=0] (mm1) to[out=-180,in=-\angleN] (u2);
      \draw[graviton] (u4) to[out=-180+\angleN,in=0] (mm2) to[out=-180,in=-\angleN] (u3);
      \draw[wl] (u2) -- (ur2) ;
      \draw[wl] (u2) -- (ul2) ;
      \draw[background] (u2) ++(\degA-90:\ldA) arc (\degA-90:-\degA-90:\ldA) ;
      \draw[wl] (u3) -- (ur3) ;
      \draw[wl] (u3) -- (ul3) ;
      \draw[background] (u3) ++(\degA-90:\ldA) arc (\degA-90:-\degA-90:\ldA) ;
      \draw[wl] (u4) -- (ur4) ;
      \draw[wl] (u4) -- (ul4) ;
      \draw[background] (u4) ++(\degA-90:\ldA) arc (\degA-90:-\degA-90:\ldA) ;
      \node[myCirclePer] () at (u2) {\icir{S}} ;
      \node[myCirclePer] () at (u3) {\icir{S}} ;
      \node[myCirclePer] () at (u4) {\icir{S}} ;
      \node[myCirclePer] () at (u3) {\icir{S}} ;
      \node[myCircleP] () at (ur2) {} ;
      \node[myCircleP] () at (ul2) {} ;
      \node[below=.36cm] () at (u2) {$n$} ;
      \node[myCircleP] () at (ur3) {} ;
      \node[myCircleP] () at (ul3) {} ;
      \node[below=.36cm] () at (u3) {$m$} ;
      \node[myCircleP] () at (ur4) {} ;
      \node[myCircleP] () at (ul4) {} ;
      \node[below=.36cm] () at (u4) {$l$} ;
    \end{tikzpicture}
    +
    \frac1{2n!m!l!}
    \begin{tikzpicture}[baseline={([yshift=-.5ex]u1)}]
      \def\upN{0} 
      \def\cN{0} 
      \def\downNa{.3} 
      \def\downNb{.6} 
      \def\laN{.5} 
      \def\lbN{.9} 
      \def\lcN{.9} 
      \def\ldN{.4} 
      \def\angleN{-90} 
      \def\leN{-1.0} 
      \def\degA{18}
      \def\lbA{.3}
      \def\lcA{-.5}
      \def\ldA{.4}
      \coordinate (u1) at (\cN,\upN) ;
      \coordinate (u2) at (\cN+\laN,\upN) ;
      \coordinate (d2) at (\cN+\laN,\upN+\leN) ;
      \coordinate (u3) at (\cN+\laN+\lbN,\upN) ;
      \coordinate (u4) at (\cN+\laN+\lbN+\lcN,\upN) ;
      \coordinate (u5) at (\cN+\laN+\lbN+\lcN+\ldN,\upN) ;
      \coordinate (mm1) at (\cN+\laN+\lbN/2,\upN+\downNa) ;
      \coordinate (mm2) at (\cN+\laN+\lbN/2+\lcN/2,\upN+\downNb) ;
      \coordinate (arrowL) at (-.2,0) ;
      \coordinate (ur2) at (\cN+\laN+\lbA,\upN+\lcA) ;
      \coordinate (ul2) at (\cN+\laN-\lbA,\upN+\lcA) ;
      \coordinate (ur3) at (\cN+\laN+\lbN+\lbA,\upN+\lcA) ;
      \coordinate (ul3) at (\cN+\laN+\lbN-\lbA,\upN+\lcA) ;
      \coordinate (ur4) at (\cN+\laN+\lbN+\lcN+\lbA,\upN+\lcA) ;
      \coordinate (ul4) at (\cN+\laN+\lbN+\lcN-\lbA,\upN+\lcA) ;
      \draw[zParticle] (u2) -- (u1) ;
      \draw[graviton] (u3) to[out=-180+\angleN,in=0] (mm1) to[out=-180,in=-\angleN] (u2);
      \draw[graviton] (u4) to[out=-180+\angleN,in=0] (mm2) to[out=-180,in=-\angleN] (u2);
      \draw[wl] (u2) -- (ur2) ;
      \draw[wl] (u2) -- (ul2) ;
      \draw[background] (u2) ++(\degA-90:\ldA) arc (\degA-90:-\degA-90:\ldA) ;
      \draw[wl] (u3) -- (ur3) ;
      \draw[wl] (u3) -- (ul3) ;
      \draw[background] (u3) ++(\degA-90:\ldA) arc (\degA-90:-\degA-90:\ldA) ;
      \draw[wl] (u4) -- (ur4) ;
      \draw[wl] (u4) -- (ul4) ;
      \draw[background] (u4) ++(\degA-90:\ldA) arc (\degA-90:-\degA-90:\ldA) ;
      \node[myCirclePer] () at (u2) {\icir{S}} ;
      \node[myCirclePer] () at (u3) {\icir{S}} ;
      \node[myCirclePer] () at (u4) {\icir{S}} ;
      \node[myCircleP] () at (ur2) {} ;
      \node[myCircleP] () at (ul2) {} ;
      \node[below=.36cm] () at (u2) {$n$} ;
      \node[myCircleP] () at (ur3) {} ;
      \node[myCircleP] () at (ul3) {} ;
      \node[below=.36cm] () at (u3) {$m$} ;
      \node[myCircleP] () at (ur4) {} ;
      \node[myCircleP] () at (ul4) {} ;
      \node[below=.36cm] () at (u4) {$l$} ;
    \end{tikzpicture}
    +
    \mO(c_{E/B}^2)
    \Bigg]
  \ .
\end{equation}
\end{widetext}
Here, the sum extends over all numbers $n$, $m$ and $l$ of superfields.
The goal will be to evaluate the right-hand-side in time domain at the background time $\bar \tau$.
Its general structure is a sum of $(j+1)$-point WQFT diagrams connected with $j$ superfields where only photons $\Del A^\mu(x)$ propagate within the diagrams.
The first term corresponds to the force (or torque) evaluated on the external EM fields and the three next terms are self-force corrections.

A generic multi-point WQFT diagram with $(j+1)$ superfield legs takes the schematic form,
\begin{equation}
  \begin{tikzpicture}[baseline={([yshift=-.5ex]u1)}]
      \def\upN{0} 
      \def\cN{0} 
      \def\downNa{-.4} 
      \def\downNb{-.4} 
      \def\laN{.5} 
      \def\lbN{.4} 
      \def\lcN{.4} 
      \def\ldN{.4} 
      \def\angleN{90} 
      \def\leN{-1.0} 
      \def\lfN{.4} 
      \def\correc{.4} 
      \def\degA{20}
      \def\lbA{.5}
      \def\lcA{.3}
      \def\ldA{.4}
      \coordinate (u1) at (\cN,\upN) ;
      \coordinate (d1) at (\cN,\upN+\leN) ;
      \coordinate (u2) at (\cN+\laN,\upN) ;
      \coordinate (ur2) at (\cN+\laN+\lbA,-\upN+\lcA) ;
      \coordinate (ul2) at (\cN+\laN+\lbA,-\upN-\lcA) ;
      \coordinate (d2) at (\cN+\laN,\upN+\leN) ;
      \coordinate (dr2) at (\cN+\laN+\lfN,\upN+\leN) ;
      \coordinate (dl2) at (\cN+\laN-\lfN,\upN+\leN) ;
      \coordinate (u3) at (\cN+\laN+\lbN,\upN) ;
      \coordinate (d3) at (\cN+\laN+\lbN,\upN+\leN) ;
      \coordinate (u4) at (\cN+\laN+\lbN+\lcN,\upN) ;
      \coordinate (d4) at (\cN+\laN+\lbN+\lcN,\upN+\leN) ;
      \coordinate (u5) at (\cN+\laN+\lbN+\lcN+\ldN,\upN) ;
      \coordinate (d5) at (\cN+\laN+\lbN+\lcN+\ldN,\upN+\leN) ;
      \coordinate (mm1) at (\cN+\laN+\lbN/2,\upN+\downNa) ;
      \coordinate (mm2) at (\cN+\laN+\lbN+\lcN/2,\upN+\downNb) ;
      \coordinate (arrowL) at (-.2,0) ;
      \coordinate (arrowH) at (0,+.2) ;
      \coordinate (u2P) at ($(u2)-(\cN+\correc,\upN)$) ;
      \draw[zParticle] (u2) -- (u1) ;
      \draw[wl] (u2) -- (ur2) ;
      \draw[wl] (u2) -- (ul2) ;
      \draw[background] (u2) ++(\degA:\ldA) arc (\degA:-\degA:\ldA) ;
      \node[right=.4cm] () at (u2) {$j$} ;
      \node[myCirclePerPer] () at (u2) {} ;
      \node[below=.05cm] () at (u1) {\ilabP{\Del Z^{\sig_0}\!(-\oma_0)\hspace{.45cm}}} ;
      \node[above=-.15cm] () at (ur2) {\ilabP{\hspace{.8cm}\Del Z^{\sig_1}\!(\oma_1)}} ;
      \node[below=-.05cm] () at (ul2) {\ilabP{\hspace{.8cm}\Del Z^{\sig_j}\!(\oma_j)}} ;
    \end{tikzpicture}
  \!\!\!\!=\!
  2\pi\del\!
  \Big(\!
    \oma_0
    -
    \sum_{i=1}^j \oma_i
    \!
    \Big)
    \mM_{\sig_0}^{\sigo...\sig_{\! j}}\!(\omao,..,\oma_j)
    \,,
    \label{exa}
\end{equation}
with amplitude $\mM$ which depends only on the frequencies and worldline background parameters.
Here, the big solid blob signifies any of the multi-point WQFT diagrams of Eq.~\eqref{exp1} where we have amputated all (incoming) superfields and external propagators.
In order to keep the discussion simple, we ignore the case of the external EM potential $A_{\rm ext}^\mu$ in the schematic form, though its inclusion is straightforward.

Let us consider the contribution of the multi-point WQFT diagram Eq.~\eqref{exa} to the regulated EOM Eq.~\eqref{exp1} in time domain evaluated at $\bar\tau$.
We thus integrate the multi-point diagram against $j$ superfield fluctuations and integrate on $\oma_0$ with a Fourier factor $\exp(-i\oma_0\bar\tau)$ at which point all frequencies become derivatives of the time domain superfields:
\begin{equation}
  \label{exa3}
    \!
    \mM_{\sig_0}^{\sigo...\sig_j}
    \!
  \Big[
    i
    \frac{\dd
    }{
    \dd\tau_1
  }
  ,..,
    i
    \frac{\dd
  }{
    \dd\tau_j}
  \Big]
  \prod_{i=1}^j
  \Del Z_{\sig_i}(\tau_i)
  \bigg|_{\tau_i\to\bar\tau}
  \ .
\end{equation}
The amplitudes $\mM$ may easily be computed and turn out to be polynomial in their arguments and finite in $d=4$.
In this case the contribution~\eqref{exa3} simply becomes a sum of $j$ superfields $\Del Z^\sig(\bar\tau)$ multiplied together and each differentiated a number (possibly zero) of times.
Crucially, since $\Del z^\sig(\bar\tau)=\Del \dot z^\sig(\bar\tau)=\Del \psi^\sig(\bar\tau)=0$, the contribution is non-zero only if each field is differentiated a minimum number of times.
Higher derivatives of $\Del Z^\sig$ are simply identical to derivatives of $Z^\sig$ itself.

We will not carry out power counting of the vertex rules explicitly but one finds that for a sufficient number $j$ of superfield legs, there are not enough differentiations to make the contribution~\eqref{exa3} non-zero.
In particular, one needs at most one incoming fluctuation in the first term of Eq.~\eqref{exp1} (i.e. $n\le1$), at most three fluctuations in the second ($n+m\le3$) and at most five fluctuations in the third and fourth ($n+m+l\le5$).

At this point we must only show that the amplitudes $\mM$ are polynomial in the frequencies and finite in $d=4$.
Non-trivial dependence on the frequencies and eventual divergencies can arise only from the loop-like integrations on the photon momenta.
The relevant integrals factorize into one-loop massive tadpoles:
\begin{equation}
  \label{integral1}
  I^{\mu_1...\mu_n}_{\rm DimReg}
  (\oma)
  =
  \int
  \dd^d k
  \frac{
    k^{\mu_1}...\,k^{\mu_n}
  }{
    (k\cdot \dot z+i\eps)^2+k_\mu k_\nu\eta_\bot^\mn
  }
  \delta(k\cdot \dot z-\omega)
  \, .
\end{equation}
Here, $k^\mu$ is the exchanged photon momentum and $\oma$ is the total energy flowing in or out of the self-interaction.
As dictated by the in-in formalism, the photon propagator is retarded with positive infinitesimal $\eps$.

The massive tadpole Eq.~\eqref{integral1} is easily computed within dimensional regularization.
Importantly, any trace $\eta_{\muo\mut} I^{\muo\mut...\mun}$ is zero because the contraction cancels the denominator and removes any scales of the integral.
With this regularization, the integral is finite and assuming all divergences to appear from self-interactions, they have thus been removed.
We can then let $d\to4$ and work in four spacetime dimensions.
The dependence on $\oma$ of the tadpole can be determined from dimensional analysis with $I^{\mu_1...\mu_n}\sim \oma^{n+1}$.

An illustrative example is given by the leading order self-force contribution where, neglecting spin and susceptibility corrections, one gets:
\begin{equation}
  \label{integral2}
  \begin{tikzpicture}[baseline={([yshift=-.5ex]u1)}]
    \def\upN{0} 
    \def\cN{0} 
    \def\downNa{.4} 
    \def\downNb{-.5} 
    \def\laN{.5} 
    \def\lbN{.8} 
    \def\lcN{.5} 
    \def\ldN{.6} 
    \def\angleN{-90} 
    \def\leN{-1.} 
    \def\correc{.4} 
    \def\degA{16}
    \def\lbA{.2}
    \def\lcA{.4}
    \def\ldA{.4}
    \coordinate (u1) at (\cN,\upN) ;
    \coordinate (u2) at (\cN+\laN,\upN) ;
    \coordinate (ur2) at (\cN+\laN+\lbA,\upN+\lcA) ;
    \coordinate (ul2) at (\cN+\laN-\lbA,\upN+\lcA) ;
    \coordinate (d2) at (\cN+\laN,\upN+\leN) ;
    \coordinate (u3) at (\cN+\laN+\lbN,\upN) ;
    \coordinate (ur3) at (\cN+\laN+\lbN+\lbA,\upN+\lcA) ;
    \coordinate (ul3) at (\cN+\laN+\lbN-\lbA,\upN+\lcA) ;
    \coordinate (u4) at (\cN+\laN+\lbN+\lcN,\upN) ;
    \coordinate (u5) at (\cN+\laN+\lbN+\lcN+\ldN,\upN) ;
    \coordinate (mm1) at (\cN+\laN+\lbN/2,\upN+\downNa) ;
    \coordinate (mm2) at (\cN+\laN+\lbN+\lcN/2,\upN+\downNb) ;
    \coordinate (arrowL) at (-.2,0) ;
    \coordinate (u2P) at ($(u2)-(\cN+\correc,\upN)$) ;
    \coordinate (uu1) at (\cN+.2,\upN) ;
    \coordinate (uu4) at (\cN-.2+\laN+\lbN+\lcN,\upN) ;
    \draw[zParticle] (u2) -- (u1) ;
    \draw[wl] (u4) -- (u3) ;
    \draw[graviton] (u3) to[out=-180+\angleN,in=0] (mm1) to[out=-180,in=-\angleN] (u2);
    \node[below] () at (uu1) {\ilabP{\Del z^\sig\!(-\omega)}} ;
    \node[below] () at (uu4) {\ilabP{\Del z^{\rho}(\omega')}} ;
    \node[myCirclePer] () at (u2) {\icir{\tau}} ;
    \node[myCirclePer] () at (u3) {\icir{\tau}} ;
  \end{tikzpicture}
  \!\!\!\!\!=
  \delta(\oma-\oma')
  \frac{q^2  \oma^3
  }{3}
  \big[
    \eta^{\sig\rho}
    \!
    -
    \dot z^\sig
    \!
    (\bar\tau)
    \dot z^{\rho}
    (\bar\tau)
    \big]
  \!
  +...
\end{equation}
When inserted in Eq.~\eqref{exa3}, the corresponding amplitude gives rise to the ALD self-force.

\sec{Self-Force Equations of Motion}
The computation of the regulated EOM Eq.~\eqref{exp1} evaluated at $\bar \tau$ may now be carried out and though there are many diagrams an automatized evaluation is easily carried out with computer algebra.
The regulated EOM results in a regulated force for the worldline trajectory and a regulated torque for the Pauli-Lubanski vector.
Because of the arbitrariness of $\bar\tau$, we simply replace it by $\tau$.

For the trajectory $z^\mu(\tau)$ we find the schematic form,
\begin{equation}\label{regulatedEOM}
  m a^\mu
  =
  f^\mu_{\rm ext}
  +
  \frac{q}{6\pi}
  \eta^\mu_{\bot\nu}
  \Big[
    q
    \dot a^\nu
  +
  f^\nu_{M}
  +
  \ce
  f_{E}^\nu
  +
  \frac{\ce q}{6\pi}
  f_{Eq}^\nu
  \Big]
  +...
\end{equation}
with $a^\mu=\ddot z^\mu$ and the ellipsis indicating terms of quadratic order in spin and susceptibility effects.
Here, the first term $f_{\rm ext}^\mu$ is the original force~\eqref{tForce} evaluated on the external EM fields and the square brackets give self-force corrections with the ALD force in the first term, spin and magnetization effects in the second and electric polarization effects in the final two terms.
For the self-force corrections we find,
\bse
  \label{sf}
\begin{align}
  f_M^\mu
  &=
  (\dot a \times \dot M)^\mu
  +\frac{\dd}{\dd\tau}
  \big[
    \dot a \times
    \big(
    M-\frac{q}{m} S
    \big)
    \big]^\mu
  \ ,
  \\
  f^\mu_{E}
  &=
  \dddot E^\mu_{\rm ext}(z)
  +
  \!\dddot a_\nu \pat^\mu\! E^\nu_{\rm ext}(z)
  +
  a^2
  \dot E^\mu_{\rm ext}(z)
  -
  \dot a^\mu
  a\!\cdot\!
  E_{\rm ext}(z)
  \nn
  \\
  &\qquad
  +
  \frac{\dd}{\dd\tau}
  \Big(
  3a^\mu
  a\cdot
  E_{\rm ext}(z)
  +
  (\dot a\times B_{\rm ext}(z))^\mu
  \Big)
  \ ,
  \\
  f_{Eq}^\mu
  &=
  \ddddot a^\mu
    +
    2\ddot a^\mu
    a^2
    +
    8
    \dot a^\mu
    \dot a\cdot a
    +
    a^\mu\frac{
    a^4
    +
    18
    a\cdot \ddot a
    +
    19
    \dot a^2
    }{2}
    \ ,
\end{align}
\ese
with the magnetic moment $M^\mu=\frac{gq}{2m} S^\mu+ \cb B_{\rm ext}^\mu(z)$.
The forces $f_M^\mu$ and $f_E^\mu$ are due to one exchange of $\Del A^\mu$ (second term of Eq.~\eqref{exp1}) and $f_{Eq}^\mu$ due to two exchanges (third and fourth terms).
Thus double radiation magnetization effects are zero at this order.
We note that the time derivatives of the cross product in the first and third lines also act on the frame (see Eq.~\eqref{crossP}).

For the torque on $S^\mu$ we find that the self-force corrections vanish at this order such that $\eta^\mu_{\bot\nu}\dot S^\nu$ is given simply by the original torque Eq.~\eqref{torque} evaluated on the external (magnetic) field.

Assuming causal boundary conditions, the regulated EOMs are, to leading order in spin and susceptibility, exact and consistent predictions of the worldline EFT framework and their validity is thus limited only by that of the EFT framework.
See also Refs.~\cite{Saketh:2021sri,Poisson:1999tv} for a discussion of the validity of the ALD equation.

The self-force results Eqs.~\eqref{sf} are to the best of our knowledge new results.
In the reviews~\cite{Teitelboim:1979px,Kosyakov:2018wek}, the case of spin is described with worldline EOMs similar to Eqs.~\eqref{tForce} and~\eqref{torque} (see e.g. Eqs.~(337) and~(338) in~\cite{Kosyakov:2018wek}) and a radiative propagator prescription is suggested as regularization but not carried out explicitly.
In fact, dimensional regularization used here is identical to this prescription as we show (and define) in the supplementary material~\cite{supp}.
This identification provides some intuition of our results: The leading order radiative magnetic field vanishes, $B_{\rm rad}^\mu=\mO(S,c_{E/B})$, which explains the vanishing of double radiation magnetization effects and leading order self-force torque effects.

Let us briefly mention the following non-trivial checks of our results with a more detailed discussion in the supplementary material~\cite{supp}.
First, our results are in complete agreement with expressions for the radiative EM fields for a generic dipole moment given in Ref.~\cite{ellis}.
Second, we have applied our methodology to the finite size coupling $a\cdot E(z)$ considered in Ref.~\cite{Galley:2010es} and reproduce the results therein except for a relative sign.
Finally, our results are consistent with the instantaneous radiative loss of four-momentum for spin given in Ref.~\cite{villarroel2}.

\sec{Outlook}
We have shown how one may systematically eliminate electromagnetic self-interactions in the worldline EFT of point-like particles deriving, in particular, novel spin and susceptibility corrections to the ALD self-force.
Straightforward generalizations and perspectives include the addition of higher order spin and finite size effects, self-force in arbitrary space-time dimensions~\cite{Galtsov:2001iv,Harte:2018iim,Galakhov:2007my,Mironov:2007nk,Cardoso:2007uy,Kazinski:2002mp} and classical non-Abelian self-interaction~\cite{Wong:1970aa,Bastianelli:2013pta,delaCruz:2020bbn,delaCruz:2021gjp,Kosyakov:2018wek,Shi:2021qsb}.

Furthermore, it would be of great interest to apply this framework to the gravitational setting where a weak-field expansion would lead to diagrams similar to the electromagnetic ones considered here except for self-interactions in the bulk giving rise to tail-effects~\cite{Galley:2015kus,Edison:2022cdu}.
A generalization to curved space would be equally exciting and allow for applications to the self-force expansion of extreme mass ratio binaries~\cite{Galley:2008ih,Adamo:2022rmp,Adamo:2023cfp,Cheung:2023lnj,Kosmopoulos:2023bwc}.

\begin{acknowledgments}
I would like to thank Alexander Broll, Gustav Mogull, Jung-Wook Kim, Raj Patil, Jan Plefka, Muddu Saketh, Jan Steinhoff and Justin Vines for very useful discussions.
I am also grateful to Raj Patil, Jan Plefka and Jan Steinhoff for comments on an earlier draft of this work.
I would also like to thank the anonymous referee for useful comments.
GUJ's research is funded by the Deutsche Forschungsgemeinschaft (DFG, German Research Foundation), Projektnummer 417533893/GRK2575 ``Rethinking Quantum Field Theory''.
\end{acknowledgments}


\appendix
\section*{Supplementary Material}
\sec{Worldline Action in General Dimensions}
\label{appendix1}
All four-dimensional Levi-Civita symbols are eliminated using the standard formula:
\begin{align}
  \eps^{\mu_1\mu_2\mu_3\mu_4}
  \eps^{\nu_1\nu_2\nu_3\nu_4}
  =
  -
  \begin{vmatrix}
    \eta^{\mu_1\nu_1}
    &
    \eta^{\mu_2\nu_1}
    &
    \eta^{\mu_3\nu_1}
    &
    \eta^{\mu_4\nu_1}
    \\
    \eta^{\mu_1\nu_2}
    &
    \eta^{\mu_2\nu_2}
    &
    \eta^{\mu_3\nu_2}
    &
    \eta^{\mu_4\nu_2}
    \\
    \eta^{\mu_1\nu_3}
    &
    \eta^{\mu_2\nu_3}
    &
    \eta^{\mu_3\nu_3}
    &
    \eta^{\mu_4\nu_3}
    \\
    \eta^{\mu_1\nu_4}
    &
    \eta^{\mu_2\nu_4}
    &
    \eta^{\mu_3\nu_4}
    &
    \eta^{\mu_4\nu_4}
  \end{vmatrix}
  \ .
\end{align}
Using this formula we find the following $d$-dimensional expression for the worldline interaction terms:
\begin{eqnarray}
  S_{\rm int}
  \!\!&=&\!\!
  -\int\dd\tau
  \Big(
  q\,\dot z
  \cdot
  A(z)
  -
    \frac{q}{m}
    \frac{
      \dot z^\sig
    F_\sig^{\ \mu}(z)
    \mS_\mu^{\ \nu}
    \dot z_\nu
    }{
      |\dot z|
    }
    +
    |\dot z|
  U
  \Big)
  \ ,
  \nn
  \\
  U
  \!\!&=&\!\!
  \frac{gq}{2m}
  \Big(
  \frac{
    F_\mn(z)\mS^\mn}{
    2}
    +
    \frac{
      \dot z^\sig
    F_\sig^{\ \mu}(z)
    \mS_\mu^{\ \nu}
    \dot z_\nu
    }{
      \dot z^2
    }
    \Big)
    \\
    &&
    -\frac{\ce+\cb}2
    \frac{\dot z^\sig
    F_\sig^{\ \mu}(z)
    F_\mu^{\ \nu}(z)
    \dot z_\nu}{
      \dot z^2}
    -\frac{\cb}4
    F_\mn(z) F^\mn(z)
  \nn
  \ .
\end{eqnarray}
Naturally, the EOMs of the main text may also be generalized to $d$ dimensions by eliminating the Levi-Civita symbols.
In that case, however, one cannot easily work with the spin vector as in Eq.~\eqref{torque} but must instead work with either the spin tensor or the Grassmann vectors.

\sec{Radiative EM Potential}
We define the radiative propagator prescription (first considered in relation to the ALD equation by Dirac in Ref.~\cite{Dirac:1938nz}) as the difference of the retarded and advanced potentials.
In analog to our dimensionally regularized integral family $I_{\rm DimReg}^{\mu_1...\mu_n}(\oma)$ from Eq.~\eqref{integral1}, we define an integral family with the radiative prescription:
\begin{align}\label{integral3}
  &I_{\rm rad}^{\mu_1...\mu_n}
  (\oma)
  =
  \frac{I_{\rm DimReg}^{\mu_1...\mu_n}(\oma)
  -
  \big(
    I_{\rm DimReg}^{\mu_1...\mu_n}(\oma)
    \big)^*}
  {2}
\end{align}
Here, the complex conjugation denoted by an asterisk turns the retarded propagator into an advanced one.

By swapping signs in the integrand of Eq.~\eqref{integral1}, $k^\mu\to-k^\mu$, $\oma\to-\oma$, one may relate the complex conjugate of the integral to itself,
\begin{align}\label{integral4}
  \big(
    I_{\rm DimReg}^{\mu_1...\mu_n}(\oma)
    \big)^*
    =
    -
    I_{\rm DimReg}^{\mu_1...\mu_n}(\oma)
    +\mO(d-4)
    \,,
\end{align}
and inserting Eq.~\eqref{integral4} into Eq.~\eqref{integral3} shows that in $d=4$ the two different prescriptions are identical: $I^{\mu_1...\mu_n}_{\rm rad}=I^{\mu_1...\mu_n}_{\rm DimReg}+\mO(d-4)$.
Note, however, that if subleading parts in $(d-4)$ would play a role, this identification would no longer hold.

An alternative way of deriving the regulated EOMs of the main text would be to split $F^\mn(z)$ into two parts,
\begin{align}\label{split}
  F^\mn(z)=F_{\rm ext}^\mn(z)+F_{\rm rad}^\mn(z)
  \ ,
\end{align}
with analogous splits for $E^\mu$ and $B^\mu$.
The first external part $F_{\rm ext}^\mn$ is the field strength of $A^\mu_{\rm ext}$ introduced in Eq.~\eqref{externalField} and the second radiative part $F_{\rm rad}^\mn$ is the field strength of the fluctuation $\Del A^\mu$ (computed either with the dimensionally regularized integrals or the identical radiative ones).
This split of the EM fields may be inserted directly into the initial EOMs Eqs.~\eqref{tForce} and~\eqref{torque} giving rise to the regulated EOMs (Eqs.~\eqref{regulatedEOM} and~\eqref{sf} for the trajectory).
As an example, for the electric polarization terms where the initial EOM is quadratic in $E^\mu$ we get three kinds of terms in the regulated EOMs with zero, one or two insertions of $E_{\rm rad}^\mu$ corresponding to $f_{\rm ext}^\mu$, $f_{\rm E}^\mu$ and $f_{\rm Eq}^\mu$ respectively.
Below, we will describe how this alternative method was used to check our results.

\sec{Radiative Field Strength of Generic Dipole} In Ref.~\cite{ellis} $F^\mn_{\rm rad}$ was computed for an arbitrary dipole moment $Q^\mn(\tau)$ of a point-like particle.
Thus, considering a point particle current,
\begin{align}
  j^\mu(x) = \int \dd \tau
  \Big(
    q \dot z^\mu
    +
    Q^\mn
    \pat_\nu
  \Big)
  \delta^4(x-z)
  \,,
\end{align}
the radiative field strength tensor as reported in Eqs.~(7) and~(24) of Ref.~\cite{ellis} read (converted to our conventions):
\begin{align}
  &F_{\rm rad}^\mn(z)=
  \frac1{6\pi}
  \delta^{[\mu}_{\alpha\vphantom{\beta}}
  \delta^{\nu]}_\beta
  \Big[
  2q
  \dddot z^{\alpha}
  \dot z^{\beta}
  -
  \dddot Q^\ab
  +4
  \dddot Q^{\alpha\sig}
  \dot z^{\beta}
  \dot z_\sig
  \nn\\ &\hspace{.4cm}
  +12
  \ddot Q^{\alpha}_{\ \sig}
  \dot z^{(\sig}
  \ddot z^{\beta)}
  +6
  \dot Q^{\alpha\sig}
  \ddot z_\sig
  \ddot z^\beta
  +8
  \dot Q^\alpha_{\ \sigma}
  \dot z^{(\sig}
  \dddot z^{\beta)}
  \nn\\ &\hspace{.4cm}
  +\ddot z^2\big(
    6
  \dot Q^{\alpha\sig}
  \dot z_\sig
  \dot z^\beta
  -
  \dot Q^{\ab}
  +6
  Q^\alpha_{\ \sig}
  \dot z^{(\sig}
  \ddot z^{\beta)}
  \big)
  +4
  Q^\alpha_{\ \sig}
  \ddot z^{(\sig}
  \dddot z^{\beta)}
  \nn\\ &\hspace{.4cm}
  +\ddot z\cdot \dddot z
  \big(6
  Q^{\alpha\sig}
  \dot z_\sig
  \dot z^\beta
  -
  Q^\ab\big)
  +2
  Q^\alpha_{\ \sig}
  \dot z^{(\sig}
  \ddddot z^{\beta)}
  \Big]
  \ .
  \label{radF}
\end{align}

This result Eq.~\eqref{radF} provides an independent check of the regulated EOMs of the main text.
Thus, the spin and susceptibility effects considered there may all be characterized by the following dipole moment,
\begin{align}\label{dipoleMoment}
  Q^\mn
  =
  2 \frac{\pat U}{\pat F_\mn}
  =
  2 \frac{\pat U}{\pat E_{[\mu}} \dot z^{\nu]}
  -\eps^{\mn\ab} \frac{\pat U}{\pat B^\alpha} \dot z_\beta
  \,,
\end{align}
with magnetic moment $\pat U/\pat B_\mu=\frac{gq}{2m} S^\mu+ \cb B^\mu(z)$ and electric moment $\pat U/\pat E_\mu=\ce E^\mu(z)$.
We have verified that upon insertion of this dipole moment in Eq.~\eqref{radF} for $F^\mn_{\rm rad}$, we reproduce the regulated EOMs of the main text from the initial EOMs by insertion of the split Eq.~\eqref{split}.
We note that since the dipole moment depends on the EM fields, one must also insert the split there.

\sec{Self-Force due to Effective Coupling $a\cdot E$}
In order to compare with the results of Ref.~\cite{Galley:2010es}, we consider self-force effects due to the finite size coupling $a\cdot E(z)$ (with $a^\mu=\ddot z^\mu$).
We define the following action,
\begin{equation}
  \label{actionC}
  S
  =
  -\int\dd\tau\Big[
    \frac{m}2
    \dot z^2
    +
    q\,
    \dot z\cdot A(z)
    +
    c\,
    \frac{a\cdot E(z)}{\sqrt{\dot z^2}}
    \Big]
  \ ,
\end{equation}
with finite size coupling $c$ and explicit time reparametrization invariance of the interaction terms in order to use proper time.

This action gives rise to the following equation of motion (with $m a^\mu = f^\mu$):
\begin{align}
  \label{EOMEa}
   f^\mu
   &=
  q E^\mu(z)
  \!+\!
  c\,
  \eta^\mn_{\bot}
  \\
  &\qquad\times
  \Big[
    (
    \pat_\nu
    +a_\nu
    )
    a\cdot E(z)
    +
    \frac{\dd}{\dd\tau}
    \big[
      a\times B(z)
    \big]_\nu
    +
    \ddot E_\nu(z)
    \Big]
  \, .
  \nn
\end{align}
Using the same methodology as in the main text we derive regularized equations of motion with,
\begin{align}
  \label{sf1}
  f^\mu = f_{\rm ext}^\mu + \frac{q^2}{6\pi} \eta_{\bot\nu}^\mu \dot a^\nu + f_{c}^\mu
  \ ,
\end{align}
where the external force $f_{\rm ext}^\mu$ is given by Eq.~\eqref{EOMEa} evaluated on the external EM fields and the second term is the ALD self-force.
The third term gives the self-force corrections due to the finite size coupling $c$ and reads:
\begin{equation}
  \label{sf2}
  f^\mu_{c}
  =
  \frac{q c}{3\pi}
  \eta^\mu_{\bot\nu}
  \Big(
    \dddot a^\mu
    +
    2\dot a^\mu a^2
    +
    6
    a^\mu a\cdot \dot a
    \Big)
  \ .
\end{equation}
This result should be compared with Eq.~(24) of Ref.~\cite{Galley:2010es}.
The two results (Eq.~\eqref{sf2} and Eq.~(24) of Ref.~\cite{Galley:2010es}) are in complete agreement except for an overall factor of $4\pi$ (due to different electromagnetic units) and a relative sign of the second term.

As with the main results of this letter, the radiative field strength tensor of Ref.~\cite{ellis} reported in Eq.~\eqref{radF} provides a strong independent check of the self-force result Eq.~\eqref{sf2}.
In the case of the finite size term $c a\cdot E$, the dipole moment reads $Q^\mn=2 c a^{[\mu} \dot z^{\nu]}$ .
Going through the same process described below Eq.~\eqref{dipoleMoment} we have independently checked the result Eq.~\eqref{sf2}.

\sec{Instantaneous Loss of Four-Momentum}
In the appendix of Ref.~\cite{villarroel2} one finds the instantaneous, radiative loss of four-momentum of a spinning, charged point-like particle which, following Ref.~\cite{villarroel2}, we denote $\dot P^\mu$.
The spinning part of the self-force $f^\mu_{M}$ is consistent with the expression for $\dot P^\mu$ given there in the sense that (up to a certain term to be discussed) they differ only by a total time derivative.
Thus, the spinning part of the self-force may written as follows,
\begin{align}\label{Prad}
  &
  \eta^\mu_{\bot\nu}
  f_{M}^\nu
  +
  \frac{6\pi}{m}
  v^\mu 
  S\cdot
  a
  \!\times\!
  E_{\rm rad}
  =
  \frac{qg}{2m}
  \big[
      v^\mu\, S\cdot a\!\times\! \dot a
      +
      (\ddot S\!\times\! a)^\mu
    \big]
    \nn
    \\[-2pt]
    &
    \qquad\quad
    +\frac{\dd}{\dd\tau}  \Big[
      \frac{qg}{2m}(a\!\times\! \dot S)^\mu+\frac{q(g-2)}{2m}(\dot a\!\times\! S)^\mu
    \Big]
\end{align}
where the right-hand-side of the first line agrees exactly with $\dot P^\mu$ given in Ref.~\cite{villarroel2}.
Apart from the correction term of the left-hand-side of the first line, this agreement is expected since the total radiated momentum is the time integral of the self-force and thus the self-force and $\dot P^\mu$ can differ only by a total time derivative which can be neglected under integration.

Let us then discuss the presence of the term proportional to $E_{\rm rad}$ in Eq.~\eqref{Prad}.
Essentially, in the regulated EOM of the main text we have cancelled a self-force contribution against an external contribution to the total force.
This is most easily understood by considering the initial force, Eq.~\eqref{tForce}, where, in the main text, we have eliminated a term which vanishes upon using the EOMs but which initially is present when one derives the EOMs from the action.
This term proportional to $S\cdot(a\times E)$ is easily seen to be zero (at leading order in spins) using the EOMs but its individual external and radiative parts do not vanish by themselves.
It is thus the radiative part of that term, which has to be added to $f^\mu_M$ in order to recover $\dot P^\mu$.


\end{document}